\documentclass[useAMS,usenatbib]{mn2e}
\usepackage{epsfig}
\usepackage{amsmath, amssymb,bm}

\title[Dead Zones around Young Stellar Objects ]{Dead Zones around
  Young Stellar Objects: Dependence on Physical Parameters }

\author[R. G. Martin et al.]{Rebecca G. Martin$^1$, Stephen
  H. Lubow$^1$, Mario Livio$^1$ and J. E. Pringle$^{1,2}$\\ $^1$Space
 Telescope Science Institute, 3700 San Martin Drive, Baltimore, MD
  21218, USA \\ $^2$Institute of Astronomy, Madingley Road,
  Cambridge, CB3 0HA, UK \\}

\begin{document}

\date{}

\pagerange{\pageref{firstpage}--\pageref{lastpage}} 
\pubyear{2011}
\maketitle

\label{firstpage}

\begin{abstract}
Angular momentum is transported outwards through an accretion disc by
magnetohydrodynamical (MHD) turbulence thus allowing material to
accrete on to the central object. The magneto-rotational instability
(MRI) requires a minimum ionisation fraction to drive turbulence in a
disc.  The inner parts of the disc around a young stellar object are
sufficiently hot to be thermally ionised. Further out, cosmic rays
ionise the surface layers and a dead zone forms at the mid-plane where
the disc is too cool for the MRI to operate.  The surface density in
the turbulent active layer is often assumed to be constant with radius
because the cosmic rays penetrate a constant layer.  However, if a
critical magnetic Reynolds number, $Re_{\rm M,crit}$, is used to
determine the extent of the dead zone, the surface density in the
layer generally increases with radius.  For small critical magnetic
Reynolds number of order 1, the constant layer approximation may be a
reasonable fit. However, MHD simulations suggest the critical magnetic
Reynolds number may be much larger, of order $10^4$. Analytical fits
for the surface density in the magnetic active layer show that
$\Sigma_{\rm m}\propto Re_{\rm M,crit}^{-2} R^{\,9/2}T^{\,2}$, at
temperature $T$ and radius $R$, are a good fit for higher critical
magnetic Reynolds number.  For the metallicity variation between our
galaxy, the LMC and the SMC, there should be no significant difference
in the extent of the dead zone. Observations suggest an increase in
the lifetime of the disc with decreasing metallicity that cannot be
explained by the dead zone structure (ignoring possible differences in
dust abundances).
\end{abstract}

\begin{keywords}
accretion, accretion discs -- planetary systems: protoplanetary discs
-- stars: pre-main-sequence
\end{keywords}

\section{Introduction}

Low-mass stars are thought to form from the free-fall collapse of a
protostellar molecular cloud core to a protostar with a disc on a
timescale of a few $10^5\,\rm yr$ \citep{shu87}. Angular momentum
transport in accretion discs is driven by turbulence thus allowing
material to accrete on to the young star. The magneto-rotational
instability (MRI) can drive turbulence if the gas is well coupled to
the magnetic field \citep{balbus91}. However, with a low ionisation
fraction the MRI is suppressed \citep{gammie96,gammie98}.  The inner
parts of a disc around a young stellar object are hot enough to be
thermally ionised. However, further out, the disc becomes layered with
an MRI turbulent (active) layer at each surface and a dead zone at the
midplane that is shielded from the ionizing radiation of cosmic rays
and X-rays from the star \citep[e.g.][]{sano00,matsumura03}.  In this
work we concentrate on circumstellar discs, but note that dead zone
formation is also favourable in circumplanetary discs
\citep{martin11a,martin11b}.

A frequently used assumption in calculations of discs with dead zones
is that the surface density of the MRI active surface layer is constant with
radius \citep[e.g.][]{armitage01,zhu09a,terquem08,matsumura09}.  The
cosmic rays enter the disc surface and are attenuated exponentially
with a stopping depth of surface density around $100\,\rm g\,cm^{-2}$
\citep{umebayashi81}.  However, this does not allow for recombination
effects that may play a significant role in determining the ionisation
of the disc. 

A more realistic way to determine the dead zone is using a magnetic
Reynolds number. MHD simulations show that magnetic turbulence cannot
be sustained if the magnetic Reynolds number is lower than some
critical value $Re_{\rm M}<Re_{\rm M,crit}$
\citep{fleming00}. However, the critical value is
uncertain. Simulations of \cite{fleming00} that compute the non-linear
outcome of instability suggest that $Re_{\rm M,crit}\approx 10^4$
without a net magnetic flux through the disc. With a magnetic flux it
could be of the order of $100$. \cite{wardle99} and \cite{balbus01}
considered the linear stability of the discs and find that the effects
of Hall electromotive forces are important and the value may be as
small as $Re_{\rm M,crit}=1$. However, conditions for linear
instability and for maintaining turbulence are not necessarily the
same \citep{balbus00}. We consider a range of values for the critical
magnetic Reynolds number from 1 to $10^4$. 

The surface layers of the disc may be ionised by cosmic rays or
X-rays.  \cite{matsumura03} find that cosmic rays determine the extent
of the dead zone under typical conditions unless the X-ray energy is
very high. We consider only cosmic ray ionisation and neglect
X-rays. Electrons in the disc are removed through dissociative
recombination with molecular ions and at a slower rate with radiative
recombination with heavy metal ions. The ionisation fraction, and
hence the extent of the dead zone, is sensitive to the number of metal
atoms because they quickly pick up the charges of molecular ions but
slowly recombine with the electrons.

The typical life time of a protostellar disc in the solar
neighbourhood, where the metallicity is $Z\approx 0.02$ is of the
order of $3-5\,\rm Myr$ \citep[e.g.][]{hernandez09}. For stars near SN
1987A in the Large Magellanic Cloud (LMC) \cite{demarchi10} find from
H$\alpha$ fluxes that the mean lifetime is of the order of $13\,\rm
Myr$ and the metallicity there is $Z\approx 0.008$.  Note however,
that these stars are likely to be slightly more massive. Similarly,
\cite{demarchi11} observed the NGC 346 cluster in the Small Magellanic
Cloud (SMC), where the metallicity is $Z\approx 0.002$, and find disc
lifetimes of around $20\,\rm Myr$. It appears from these results that
the life time of the disc increases with decreasing metallicity.  Dead
zones affect the accretion rate through the disc and thus lifetime of
disc accretion and so we consider how the dead zone varies with
metallicity.

In Section~\ref{sec:layer} we describe the layered disc model where
the extent of the dead zone is determined with the critical magnetic
Reynolds number. We consider thermal ionisation, cosmic ray ionisation
and recombination. In Section~\ref{disc} we choose a fiducial disc
model and find the dead zone and resulting accretion rate through the
disc for varying metallicity and critical magnetic Reynolds number. We
investigate the extent to which assuming a constant surface density in
the active layer can approximate this and how sensitive the dead zone
is to the metallicity. In Section~\ref{an} we find analytical
approximations to the surface density in the active layer of the disc.

\section{Layered Disc Model}
\label{sec:layer}

In an accretion disc \citep{lyndenbell74,pringle81} around a central
object of mass $M$, where the material is in Keplerian orbits with
angular velocity $\Omega=\sqrt{GM/R^3}$ at radius $R$, the kinematic
turbulent viscosity may be parametrised by the $\alpha$-prescription
\begin{equation}
\nu=\alpha \frac{c_{\rm s}^2}{\Omega}
\label{visc}
\end{equation}
\citep{shakura73}, where $\alpha$ is a dimensionless parameter that is
not well constrained.  Numerical simulations of MHD turbulence find
$\alpha\gtrsim 0.01$ \citep{brandenburg95,stone96, fromang07,guan09,
  davis09}. Observational evidence from dwarf nova and X-ray outbursts
suggests $\alpha\sim 0.1-0.4$ \citep{king07}.  In this work we take
$\alpha=0.1$ but discuss implications of a lower value. The sound
speed is given by
\begin{equation}
 c_{\rm s}=\sqrt{\frac{{\cal R }T}{\mu }},
\end{equation}
where $T$ is the midplane temperature of the disc, $\cal R$ is the
gas constant,  $\mu=2.3$ is the mean particle weight.  The
disc is viscous only in the turbulent active surface layer of the
disc, not in the dead zone.

The dead zone is determined by where the MRI is unsustained because of
poor coupling of the magnetic field to the disc. If turbulence is
damped on a scale of $\lambda$, then the growth rate of the MRI is
$V_{\rm A}/\lambda$, where $V_{\rm A}=B/(4 \pi \rho)^\frac{1}{2}$ is
the Alv\'{e}n speed, $B$ is the magnetic field and $\rho$ is the
density. \cite{hawley95} use numerical simulations to find the shear
stress due to turbulence induced by the instability to be
$w_{r\phi}\approx \rho V_{\rm A}^2$. In terms of the \cite{shakura73}
$\alpha$-parameter this is $w_{r\phi}=\alpha \rho c_{\rm s}^2$ and so
equating these we find $V_{\rm A}\approx \alpha^\frac{1}{2}c_{\rm
  s}$. The growth rate of the Ohmic diffusion is $\eta/\lambda^2$,
where $\eta$ is the diffusivity of the magnetic field. Complete
damping occurs when all turbulence scales less than the scale height,
$H=c_{\rm s}/\Omega$, are damped. This means that $\lambda < H$, or
equivalently we can define the magnetic Reynolds number
\begin{equation}
Re_{\rm M}=\frac{\alpha^\frac{1}{2}c_{\rm s}H}{\eta} ,
\label{rey}
\end{equation}
and the disc is dead if $Re_{\rm M}(R,z)<Re_{\rm M,crit}$. Note
that several definitions of the magnetic Reynolds number have been
used.  For example, \cite{fleming00} defined it as above but without
the factor of $\alpha^\frac{1}{2}$.  We consider a range of values for
$Re_{\rm M,crit}$. The Ohmic resistivity is
\begin{equation}
\eta=\frac{234 \,T^\frac{1}{2}}{x_{\rm e}}\,\rm cm^2\,s^{-1}
\end{equation}
\citep{blaes94}, where the electron fraction is
\begin{equation}
x_{\rm e}=\frac{n_{\rm e}}{n_{\rm n}},
\end{equation}
where $n_{\rm e}$ is the number density of electrons and $n_{\rm n}$
is the total number density.  With vertical hydrostatic equilibrium
the number density is
\begin{equation}
n_{\rm n}=n_{\rm c}\exp \left[-\frac{1}{2}\left(\frac{z}{H}\right)^2\right],
\end{equation}
where $z$ is the height above the mid-plane of the disc.  The number
density at the midplane of the disc, where $z=0$, is
\begin{equation}
n_{\rm c}=\frac{\Sigma}{\sqrt{2 \pi}\mu m_{\rm H} H},
\end{equation}
where $m_{\rm H}$ is the mass of hydrogen.  In the following two
sections we consider the electron fraction in the disc with thermal
ionisation, cosmic ray ionisation and recombination.

\subsection{Thermal Ionisation}
\label{thermal}

The dominant thermal ionisation ions in protostellar discs are Na$^+$
and K$^+$ \citep{umebayashi88}.  However the K$^+$ ion is more
important at the onset of dynamically interesting ionisation levels
because of its smaller ionisation potential. The Saha equation for the
electron fraction from thermal ionisation can be approximated by
\begin{align}
x_{\rm e} &  =  6.47\times 10^{-13}\,\,\left(10^{[K/H]}\right)^\frac{1}{2}\left(\frac{T}{10^3 {\,\rm K}}\right)^\frac{3}{4}\cr
 & \,\,\,\, \times \left(\frac{2.4\times 10^{15}}{n_{\rm n}}\right)^\frac{1}{2}\frac{\exp(-25188/T)}{1.15\times 10^{-11}}
\label{xet}
\end{align}
\citep{balbus00}, where $[K/H]=\log_{10}(K/H)-\log_{10}(K/H)_{\rm
  solar}$ is the potassium abundance relative to hydrogen and
$\log_{10}(K/H)_{\rm solar}=-7$. We consider the effect of varying
$[K/H]$ in Sections~\ref{a}. Thermal ionisation is most important in
the inner parts of the disc where the temperatures are higher and it
falls off exponentially.

\subsection{Recombination}
\label{re}

When thermal ionisation becomes negligible, the ionisation fraction is
found with the balance of cosmic ray ionisation and recombination
effects. Electrons may be captured by dissociative recombination with
molecular ions (of density $n_{\rm M^+}$) and radiative recombination
with heavy-metal ions (of density $n_{\rm m^+}$). Charge will be
transferred from molecular ions to metal atoms and so the electron
fraction depends also upon the metal fraction
\begin{equation}
x_{\rm m}=\frac{n_{\rm m}}{n_{\rm n}},
\end{equation}
where $n_{\rm m}$ is the number density of metals.  The standard
notation for metallicity is by mass fraction, $Z$. This is related to
the metal fraction with
\begin{equation}
x_{\rm m}=\frac{Z}{\mu}
\label{mf}
\end{equation}
where $\mu$ is the average mass of a particle in units of the mass of
hydrogen. For example, the molecule of highest mass taking part in
the reactions may be CO, which has a mass of 28 units. In our galaxy
$Z=0.02$ so $x_{\rm m}\gtrsim 7\times 10^{-4}$. In the LMC
$Z=0.008$ is equivalent to $x_{\rm m}\gtrsim 3\times 10^{-4}$ and in
the SMC $Z=0.002$ is equivalent to $x_{\rm m}\gtrsim 7\times 10^{-5}$.

Assuming the rates are the same for all species, the rate equation for
the electron density is
\begin{equation}
\frac{dn_{\rm e}}{dt}=\zeta n_{\rm n}-(\beta_{\rm d}n_{\rm M^+}+\beta_{\rm r} n_{\rm m^+})n_{\rm e}
\label{ne}
\end{equation}
and for the molecular ion density
\begin{equation}
\frac{d(n_{\rm M^+})}{dt}=\zeta n_{\rm n} -(\beta_{\rm d} n_{\rm e}+\beta_{\rm t} n_{\rm m})n_{\rm M^+}.
\label{nions}
\end{equation}
The ionisation rate, $\zeta$, is discussed in the next section.  The
recombination rate coefficients are
\begin{equation}
\beta_{\rm r}=3\times 10^{-11}T^{-\frac{1}{2}}\,\rm cm^3\,s^{-1}
\end{equation}
for the radiative recombination of electrons with metal ions,
\begin{equation}
\beta_{\rm d}=3\times 10^{-6}T^{-\frac{1}{2}}\,\rm cm^3\,s^{-1}
\end{equation}
for the dissociative recombination of electrons with molecular
ions and
\begin{equation}
\beta_{\rm t}=3\times 10^{-9} \,\rm cm^3\,s^{-1}
\end{equation}
for the charge transfer from molecular ions to metal atoms. 
Conservation of charge tells us that
\begin{equation}
n_{\rm e}=n_{\rm M^+}+n_{\rm m^+}.
\label{charge}
\end{equation}
Solving equations~(\ref{ne}) and~(\ref{nions}) in steady state with
equation~(\ref{charge}) gives the cubic equation 
\begin{equation}
x_{\rm e}^3+\frac{\beta_{\rm t}}{\beta_{\rm d}}x_{\rm m}x_{\rm e}^2-\frac{\zeta}{\beta_{\rm d} n_{\rm n}}x_{\rm e}-\frac{\zeta \beta_{\rm t}}{\beta_{\rm d} \beta_{\rm r}n_{\rm n}}x_{\rm m}=0
\label{cubic}
\end{equation}
\citep{oppenheimer74,spitzer98,millar97} that can be solved to find
the electron density. There are three solutions to
equation~(\ref{cubic}) but only one that is both real and positive.

The solution to equation~(\ref{cubic}) can be simplified for the case
where there is zero metallicity, $x_{\rm m}=0$, and find
\begin{equation}
x_{\rm e}=\sqrt{\frac{\zeta}{\beta_{\rm d} n_{\rm n}}}
\label{low}
\end{equation}
and also when the metals dominate so that $x_{\rm m}\gg x_{\rm e}$ and
\begin{equation}
x_{\rm e}=\sqrt{\frac{\zeta}{\beta_{\rm r} n_{\rm n}}}
\label{high}
\end{equation}
\citep[e.g.][]{matsumura03}.  We use these limits in
Section~\ref{disc}. We note that this model does not take the effects
of dust into account that may increase the size of the dead zone
\citep[e.g.][]{turner08,okuzumi11}.

\subsection{Cosmic Ray Ionisation}

We consider the only external ionisation source to be
cosmic rays with ionisation rate
\begin{equation}
\zeta=\zeta_0 \exp\left(-\frac{\Sigma_{\rm m}/2}{ \chi_{\rm cr}}\right),
\label{ion2}
\end{equation}
\cite[where we have neglected the second term in the equation
  by][]{sano00}, $\zeta_0=10^{-17}\,\rm s^{-1}$ \citep{spitzer68} and
$\chi_{\rm cr}=100\,\rm g\,cm^{-2}$ \citep{umebayashi81}.  The total
surface density in the active layers is
\begin{equation}
\Sigma_{\rm m}=2 \mu m_{\rm H} \int_{z_{\rm crit}}^\infty n_{\rm n}\,dz = \Sigma\, {\rm erfc} \left(\frac{z_{\rm crit}}{\sqrt{2} H}  \right),
\label{sa}
\end{equation}
where $z_{\rm crit}$ is the height of the disc above the mid-plane
where the dead zone ends which we find in Section~\ref{dz}.

\subsection{X-ray Ionisation}

Young stellar objects may be active X-ray sources
\citep[e.g][]{koyama94}. However, \cite{matsumura03} find that cosmic
ray ionisation dominates X-ray ionisation. For example, comparing
their figures~2 and~3 the dead zone is much larger in both height and
radius with only X-rays.  X-rays may dominate only if the energy is
very high, $k T_{\rm X}=5-10\,\rm keV$, but this is much higher than
for most observed sources. In this work we neglect X-ray ionisation
and concentrate on the effects of cosmic ray ionisation.

\subsection{Dead Zone}
\label{dz}

We have found the electron fraction with thermal ionisation and cosmic
ray ionisation with recombination. In order to determine the dead zone
height we take the electron fraction to be the maximum of
equation~(\ref{xet}) and the solution to equation~(\ref{cubic}).  Now
we can find the magnetic Reynolds number in equation~(\ref{rey}) and
for each radius, $R$, find the critical height in the disc, $z_{\rm
  crit}$ by solving $Re_{\rm M}=Re_{\rm M,crit}$. Where thermal
ionisation is dominant, this is straightforward. However, it
is slightly more complicated where cosmic ray ionisation
dominates. For example, with the high metallicity limit for the
electron fraction (equation~\ref{high}) we solve
\begin{align}
-\frac{1}{2}\left(\frac{z_{\rm crit}}{H}\right)^2= 
& \ln\left(\frac{\alpha^\frac{1}{2}c_{\rm
    s}H\zeta_0^\frac{1}{2}}{234\, T^\frac{1}{2}\beta_{\rm r}^\frac{1}{2} Re_{\rm M,crit}n_{\rm c}^\frac{1}{2}}\right)^2 \cr
& -\frac{\Sigma/2}{ \chi_{\rm cr}}{\rm erf}\left(\frac{z_{\rm crit}}{\sqrt{2}H}\right).
\label{eq}
\end{align}
In the limit of low metallicity we replace $\beta_{\rm r}$ with $\beta_{\rm
  d}$.  Given a total surface density and temperature, this equation
can be solved numerically to find $z_{\rm crit}$ and hence the active
layer surface density given in equation~(\ref{sa}) and the dead zone
surface density $\Sigma_{\rm d}=\Sigma-\Sigma_{\rm m}$. We illustrate
this with an example in the next section.

\section{Disc Properties}
\label{disc}

We choose a fiducial model for the disc structure around and star of
mass $M=1\,\rm M_\odot$ and consider properties of the dead zone. We
take the total surface density to be
\begin{equation}
\Sigma=\Sigma_0\,\left(\frac{R}{ \rm AU}\right)^{-s},
\label{surf}
\end{equation}
where $\Sigma_0=10^3\,\rm g\,cm^{-2}$ and the midplane temperature is
\begin{equation}
T=T_0\,\left(\frac{R}{\rm AU}\right)^{-t},
\label{temp}
\end{equation}
where $T_0=100\,\rm K$.  We choose the disc such that without the dead
zone, it has a constant accretion rate at all radii. This requires
$\dot M \propto \nu \Sigma =\,\rm const$ and so we take
$s+t=1.5$. Unless otherwise stated, we choose $s=t=0.75$.  This disc
is in a steady state if it is supplied with a constant source of mass
in the outer parts and the disc is fully MRI active.  However, we now
show that the disc may have a dead zone.

\subsection{Active Layer}
\label{a}

\begin{figure}
\includegraphics[width=8.8cm]{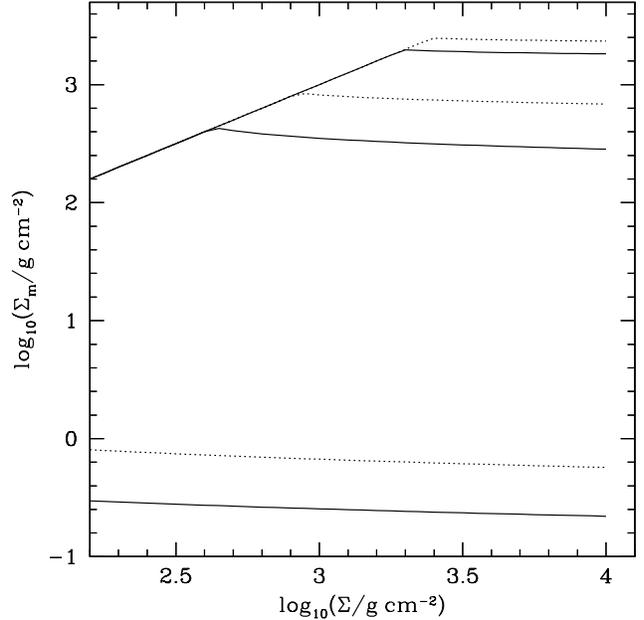}
\caption{For a fixed radius in the disc of $R=1\,\rm AU$, the surface
  density of the active layer in the disc is shown as a function of
  the total surface density. The upper two lines show $Re_{\rm
    M,crit}=1$, the middle two lines show $Re_{\rm M,crit}=100$ and
  the lower two $Re_{\rm M,crit}=10^4$. The dotted lines show the high
  metallicity limit for the electron fraction found in
  equation~(\ref{high}) and the solid lines have $x_{\rm m}=10^{-12}$
  (upper solid line), $x_{\rm m}=10^{-10}$ (middle solid line) and
  $x_{\rm m}=10^{-7}$ (lower solid line) found by solving the
  cubic~(\ref{cubic}) for the electron fraction.}
\label{thick}
\end{figure}

In Fig.~\ref{thick} we show, at a fixed radius in the disc, $R=1\,\rm
AU$, the surface density in the active layer as a function of the
total surface density. We vary the metallicity and critical magnetic
Reynolds number in the disc. The dotted lines show the high
metallicity limit for the electron fraction given in
equation~(\ref{high}). The solid lines show intermediate metallicities
found by solving the cubic equation~(\ref{cubic}) for the electron
fraction.  For fixed radius, the active layer reaches a constant
surface density as the total surface density increases. The active
layer is smallest with a low metallicity and a high critical magnetic
Reynolds number.

However, the finite metallicities considered in Fig.~\ref{thick} are
tiny.  Even the highest finite metal fraction we consider here,
$x_{\rm m}=10^{-7}$, corresponds to a very small metallicity of about
$Z=3\times 10^{-6}$ (see equation~\ref{mf}) that is several orders of
magnitude smaller than that in the SMC.  The metallicity required for
the dead zone to be significantly different from the high metallicity
limit is very small, unless the critical magnetic Reynolds number is
found to be significantly larger than $10^4$. Hence, for reasonable
metallicities, those in the observable universe, the dead zone
structure does not depend on metallicity and the high metalliticy
limit is appropriate.  In the rest of this work we consider only the
high metallicity limit.

\begin{figure}
\includegraphics[width=8.9cm]{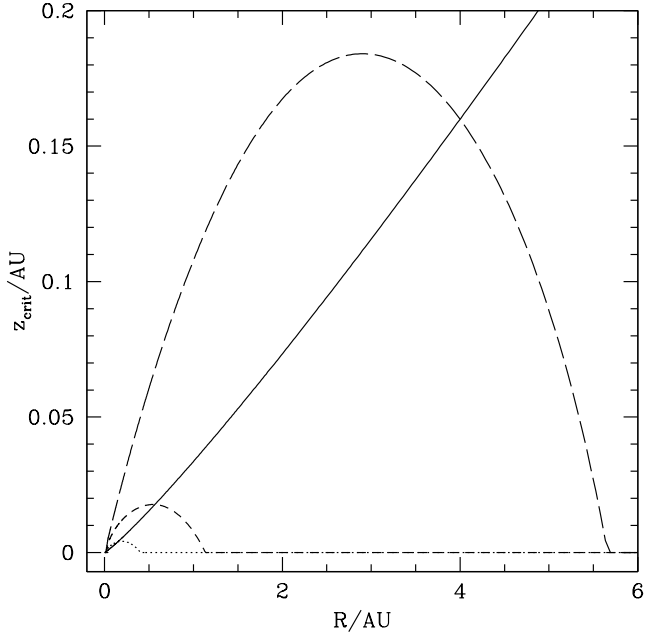}
\caption{The height of the dead zone as a function of radius for
  critical magnetic Reynolds number of $1$ (dotted line), $100$
  (short-dashed line) and $10^4$ (long-dashed line) in the high
  metallicity limit. The solid line shows the scale height of the
  disc, $H$. Note that the axis scales are not the same.}
\label{zones}
\end{figure}

\begin{figure}
\includegraphics[width=8.9cm]{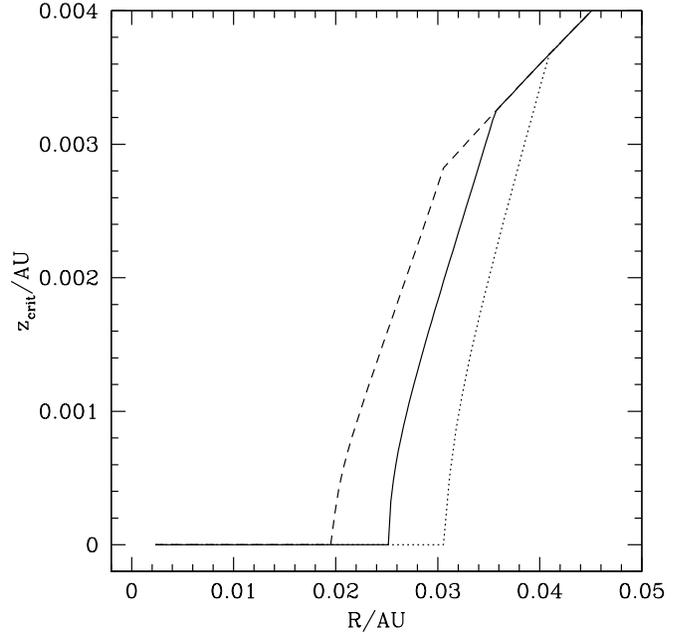}
\caption{The height of the inner parts of the dead zone as a function
  of radius for a critical magnetic Reynolds number of $100$ in each
  case with solar abundance, $[K/H]=0$, (solid line, as used for the
  rest of the work), $[K/H]=2$ (dotted line) and $[K/H]=-2$ (dashed
  line) in the high metallicity limit. The three dead zone heights are
  different where the electron fraction due to thermal ionisation
  (equation~\ref{xet}) is larger than that due to the cosmic ray
  ionisation (equation~\ref{high}). At larger radii thermal ionisation
  becomes negligible and the dead zone is found with the cosmic ray
  ionisation that does not depend on the parameter $a$. Note that the
  scales of the axes are not the same.}
\label{zones2}
\end{figure}

In Fig.~\ref{zones} we show the height of the dead zone as a function
of radius in the disc for varying critical magnetic Reynolds number in
the high metallicity limit. As the critical magnetic Reynolds number
increases, the size of the dead zone increases.  The results are
similar to those of \cite{matsumura03} who find the dead zone in a
disc with a different surface density and temperature distribution. In
Fig.~\ref{zones2} we also plot the inner most parts of the dead zone
and see that the effect of varying the constant $[K/H]$ (the Potassium
abundance relative to hydrogen, see Section~\ref{thermal}) is not very
significant. It only affects the innermost parts of the disc where
thermal ionisation is the dominant source. Further out where cosmic
ray ionisation becomes dominant the dead zone does not depend on
$[K/H]$. In the rest of the work we take the solar abundance value,
$[K/H]=0$.

\begin{figure}
\includegraphics[width=8.9cm]{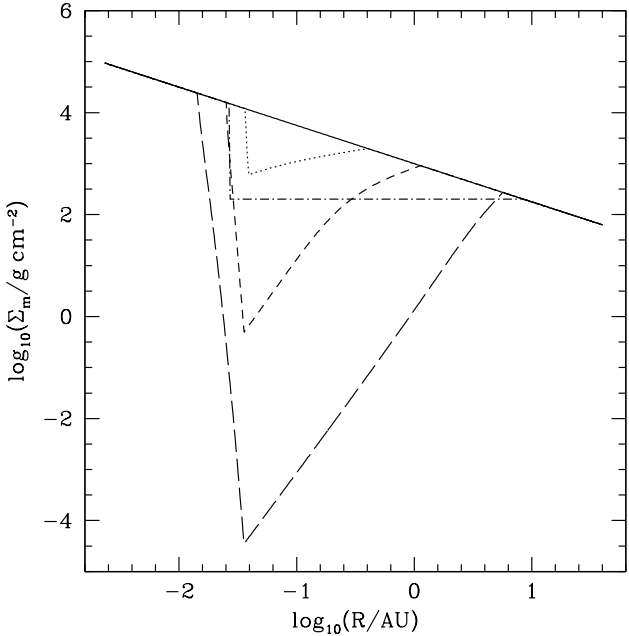}
\caption{The total surface density (solid line) and the surface
  density in the active layer as a function of radius in the disc for
  critical magnetic Reynolds number of $1$ (dotted line), $100$
  (short-dashed line) and $10^4$ (long-dashed line) in the high
  metallicity limit. The inner parts of the disc are thermally ionised
  so $\Sigma_{\rm m}=\Sigma$ and similarly the outer parts of the disc
  are fully ionised by cosmic rays. The dot-dashed line shows the
  surface density in the active layer for the constant layer
  assumption where the disc is dead if $T<T_{\rm crit}=1500\,\rm K$
  and $\Sigma > \Sigma_{\rm crit}=200\,\rm g\,cm^{-2}$.  }
\label{layer}
\end{figure}

In Fig.~\ref{layer} we show the surface density of the active layer as
a function of radius of the disc for varying critical magnetic
Reynolds number in the high metallicity limit. The inner parts of the
disc are thermally ionised and the outer parts are fully ionised by
cosmic rays. Hence in these regions, the surface density of the active
layer is equal to the total surface density, $\Sigma_{\rm m}=\Sigma$.

In the region where a dead zone exists, the surface density in the
active layer initially decreases sharply. This is because the
temperature in the disc decreases with radius ($T\propto R^{-3/4}$)
and the thermal ionisation decreases exponentially with
temperature. However, once the ionisation through cosmic rays
dominates the thermal ionisation, the active layer surface density
increases with radius until the whole disc is turbulent.
Independently, the decreasing temperature distribution has the effect
of lowering the active layer surface density in this region. However,
the dominant effect is the scale height of the disc, $H$, that
increases with radius ($H\propto R^{9/8}$).  As described in
Section~\ref{sec:layer}, the turbulence is completely damped when it
is damped on all scales $<H$. The damping decreases with radius
because of the increasing scale height and thus the active layer
surface density increases.

The constant surface density in the active layer that is more
generally assumed is also plotted in the dot-dashed line in
Fig.~\ref{layer}. We take as an example a disc which is dead if
$T<T_{\rm crit}=1500\,\rm K$ and $\Sigma>\Sigma_{\rm crit}=200\,\rm
g\,cm^{-2}$. The active surface density determined by the critical
magnetic Reynolds number is very different from a constant except for
a the low critical magnetic Reynolds number. For $Re_{\rm M,crit}=1$,
the surface density of the active layer is close to constant but much
larger than the generally assumed values that are less than around
$200\,\rm g\,cm^{-2}$. The minimum surface density in the active
layer, that occurs close to the inner edge of the dead zone, can be
very small especially for high critical magnetic Reynolds number. This
is not well represented by the assumed constant value.

\subsection{Accretion Rate}
\label{ar}

\begin{figure}
\includegraphics[width=8.9cm]{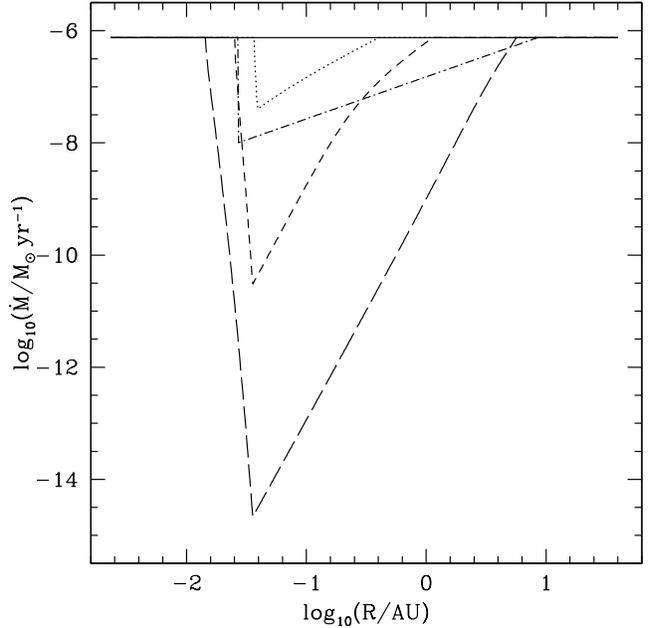}
\caption{The accretion rate through the active layer of the disc in
  the high metallicity limit. The solid line shows a disc that is
  fully turbulent without a dead zone. The other three lines have a
  dead zone with a critical magnetic Reynolds number of $1$ (dotted
  line), $100$ (short-dashed line) and $10^4$ (long-dashed
  line). These discs cannot be in steady state. The dot-dashed line
  shows the accretion rate through a disc with the constant surface
  density in the active layer (as in Fig.~\ref{layer}).}
\label{mdot}
\end{figure}

The accretion rate through the active layer of the disc is
approximately given by
\begin{equation}
\dot M=3 \pi \nu \Sigma_{\rm m}.
\end{equation}
In Fig.~\ref{mdot} we show the accretion rate as a function of radius
in the disc for our fiducial model for varying critical magnetic
Reynolds number in the high metallicity limit. The thin active layer
severely limits the accretion rate over the dead zone by up to several
orders of magnitude compared with that through the outer turbulent
parts of the disc.  These discs cannot be in steady state.

The mass in the dead zone must increase in time because it is
continually supplied by the active layer.  With sufficient accretion
on to the disc, the outer parts of the dead zone can become
gravitationally unstable.  The extra heating from the viscosity due to
self gravity can trigger the MRI and with the sudden increase in the
turbulence a significant fraction of the disc is accreted in an
outburst.  After the outburst the disc cools, the dead zone reforms
and the cycle repeats \citep{armitage01,zhu09b}. This outburst
mechanism, known as the gravo-magneto instability, should be similar
no matter how the extent of the dead zone is determined.  The
outbursts can be explained by plotting the accretion rate through the
disc against the surface density at a fixed radius. There are two
possible steady disc solutions, one that is fully turbulent and a
second that is self gravitating. Outbursts occur when there is no
steady solution for a given accretion rate at a specific radius. The
disc moves between the two solutions in a limit cycle
\citep{martin11c}.

The typical bolometric luminosity of protostars is smaller than would
be expected from the infall rate and duration of the protostellar
phase \citep{kenyon90}. Observations of T~Tauri stars suggest that at
an age of around $1\,\rm Myr$ the accretion rate through the disc is
around $10^{-8}\,\rm M_\odot\,yr^{-1}$
\citep[e.g.][]{valenti93,hartmann98}.  One solution to this luminosity
problem is thought to be time-dependent accretion on to the star that
may be caused by dead zone formation in the disc. The disc spends the
majority of its time with a dead zone and a restricted accretion rate
on to the star. A low critical magnetic Reynolds number of around one
could help to explain the observed accretion rates (see
Fig.~\ref{mdot}). In a time dependent disc, the accretion on to the
central star will be of the order of the minimum accretion rate
through the disc after a viscous timescale,
\begin{equation}
t_{\nu} =\frac{R^2}{\nu},
\end{equation}
at that radius. For example, at a radius of $0.1\,\rm AU$, the viscous
timescale is short at around $450\,\rm yr$.  It is difficult to
reconcile a model with a high critical magnetic Reynolds number to the
observations of T~Tauri star accretion rates unless there is a source
of external heating on the disc.

\section{Analytic Approximations}
\label{an}

Fig.~\ref{layer} suggests that the surface density in the active layer
above the dead zone can be approximated by a power law in radius for
high critical magnetic Reynolds numbers.  When the surface density is
sufficiently large, the active layer surface density is insensitive to
the total surface density (as shown in Fig.~\ref{thick}). This,
however, does not translate into the active layer surface density
being constant in radius. As described in Section~\ref{a}, the
dominant effect is the scale height of the disc, $H$, that increases
with radius leading to an increase in the active layer surface
density.

\begin{figure}
\includegraphics[width=8.9cm]{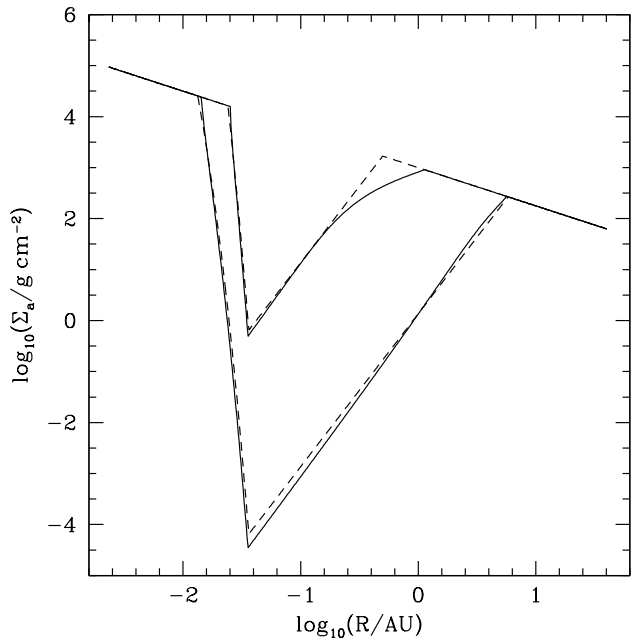}
\caption{The surface density in the active layer. The solid lines show
  numerical solutions. The dashed lines show analytic
  approximations. The upper lines have $Re_{\rm M,crit}=100$ and the
  lower lines have $Re_{\rm M,crit}=10^4$.}
\label{analytic}
\end{figure}

An analytic approximation to the active layer surface density is found
by neglecting the third term in equation~(\ref{eq}) and then
approximating $\Sigma_{\rm m}\propto \exp [-(z_{\rm
    crit}/H)^2/2]$. For cosmic ray ionisation
\begin{align}
\Sigma_{\rm a,cr}= & \,\,1.36\,\left(\frac{\alpha}{0.1}\right)\left(\frac{T}{100}\right)^{2}\left(\frac{Re_{\rm M,crit}}{10^4}\right)^{-2}
\left(\frac{M}{\rm M_\odot}\right)^{-\frac{3}{2}}
\cr
& \times \left(\frac{\zeta_0}{10^{-17}\,\rm s^{-1}}\right)\left(\frac{R}{1\,\rm AU}\right)^{9/2}\,\rm g\,cm^{-2}.
\label{fit}
\end{align}
In regions where thermal ionisation is the dominant source, the layer
can be approximated with
\begin{align}
\Sigma_{\rm a,t} = &\,\,
0.2\,\left(\frac{\alpha}{0.1}\right)\left(\frac{T}{1500}\right)^{3}\left(\frac{Re_{\rm
    M,crit}}{10^4}\right)^{-2}
\left(\frac{R}{0.03\,\rm AU}\right)^{9/2}\cr & \times
\left(\frac{M}{\rm M_\odot}\right)^{-\frac{3}{2}}
\exp\left[-25188\left(\frac{1}{T}-\frac{1}{1500}\right)\right] \,\rm g\,cm^{-2}.
\label{fit2}
\end{align}
Now we can fit the whole active layer surface density analytically. In
Fig.~\ref{analytic} we show the analytic approximation to the surface
density in the active layer  is a good fit for high critical
magnetic Reynolds number, $Re_{\rm M,crit} \gtrsim 100$. Note that
even though we have only shown one temperature and surface density
distribution in this work, these fits are valid for any distribution.

The accretion rate through the layer can be approximated by
\begin{align}
\dot M = &  1.0\times 10^{-9} \left(\frac{\alpha}{0.1}\right)^{2} \left(\frac{T}{100}\right)^{3}\left(\frac{Re_{\rm M,crit}}{10^4}\right)^{-2}\cr
& \times \left(\frac{M}{\rm M_\odot}\right)^{-2}\left(\frac{\zeta_0}{10^{-17}\,\rm s^{-1}}\right)\left(\frac{R}{1\,\rm AU}\right)^{6}\rm M_\odot\,\rm yr^{-1}
\label{mdoteq}
\end{align}
in regions where cosmic ray ionisation is the dominant source. The
minimum accretion rate through the disc occurs where cosmic ray
ionisation takes over from thermal ionisation. By finding where the
surface densities in equations~(\ref{fit}) and~(\ref{fit2}) are equal,
we find this occurs at a temperature of $T\approx 10^3\,\rm K$. The
radius corresponding to this temperature has the smallest accretion
rate (see Fig.~\ref{mdot}). As explained in Section~\ref{ar}, the
accretion on to the central object will be limited by this minimum
over a viscous timescale at that radius.

If the viscosity $\alpha$ parameter is smaller than the value we have
taken in this work, of 0.1, then both the active layer surface density
and accretion rate would be smaller than predicted here (see
equations~\ref{fit} and~\ref{fit2}). The disc would become even more unstable to the
gravo-magneto instability. The results shown here represent an upper
limit to the surface density in the active layer.

\section{Discussion}
\label{concs}

There are three non-ideal MHD effects in the generalised Ohm's law,
Ohmic resistivity, Hall effect and ambipolar diffusion, that dominate
for different regimes. The Ohmic term, that we have used here,
dominates at high density and very low ionisation. However, the
ambipolar diffusion dominates for low density and high ionisation and
the Hall effect dominates in a region between these two extremes
\citep{wardle97}.  Recent work suggests that the MRI in the surface
layers of discs may be dictated by ambipolar diffusion
\citep{perezbecker11a,perezbecker11b,bai11} or the Hall effect
\citep{wardle11} rather than Ohmic resistivity as we have assumed
here.  \cite{wardle11} applied a model with all three effects to the
minimum mass solar nebula and found that at a radius of $1\,\rm AU$
there is around an order of magnitude increase or decrease in the
active layer surface density, depending on whether the field is
parallel or antiparallel to the rotation axis, respectively, compared
with a disc with only Ohmic resistivity. These extra effects could
change the surface density in the active layer and the accretion rate
through the disc predicted here by around an order of
magnitude. However, Ohmic resistivity has the advantage of being
independent of the magnetic field \citep{fleming00} and the values it
produces for the surface density of the active layer represent an
average for the values obtained over a range of vertical magnetic
fields \citep[e.g.][]{wardle11}.

\cite{fromang02} use a similar approach to the work presented here but
find that the active layer in the disc decreases with radius such that
the dead zone surface density is approximately constant.  The
temperature distribution they use, shown in their Figure~1, has a very
steep radial dependence in the region where the dead zone forms.  With
the analytical fits given here, it is clear that the constant active
layer surface density can be recovered if the temperature of the disc
has a radius dependence $T\propto R^{-9/4}$. If the power is even
smaller, $<-9/4$, then the active layer will decrease with radius as
found by \cite{fromang02}. The analytical fits presented here in
Section~\ref{an} would provide an approximation to the disc in this case
too because they can describe any surface density and temperature
distribution.

If cosmic rays are present, the depth of the active layer is
insensitive to the presence of X-rays because they do not penetrate so
far. However, X-rays could still increase the ionisation fraction, and
hence the associated effective $\alpha$, in the active
layer. \cite{igea99} showed that the X-ray ionisation rate near the
surface is significantly larger than that of cosmic rays. The
accretion rate through the layer may be higher than predicted in
equation~(\ref{mdoteq}) if this is taken into account and perhaps
could help to explain the T~Tauri accretion rates for higher critical
magnetic Reynolds number.

Observed FU Orionis systems may be one manifestation of the
time-dependent accretion phenomenon. These luminous young stellar
objects are found in star forming regions \citep{hartmann96}. Their
optical brightness increases by five magnitudes or more on a timescale
of years and decays on a timescale of 50-100 years \citep{herbig77}.
The timescale and brightness of accretion outbursts from the
gravo-magneto instability are similar to the observed FU Orionis
outbursts \citep{armitage01,zhu09b}. Time-dependent numerical
simulations are needed to investigate these effects more fully in a
disc with a dead zone determined by the critical magnetic Reynolds
number. By comparing the resulting outbursts with FU Orionis
observations the value of the critical magnetic Reynolds number may be
further constrained.

\section{Conclusions}

A typical assumption used in time-dependent simulations of accretion
discs with dead zones is that the surface density in the MRI active
layer above the dead zone is constant with radius. However, when the
dead zone is identified by the value of the critical magnetic Reynolds
number including the effects of recombination, the active layer
surface density is generally found to increase with radius. The
constant layer is best reproduced with a low critical magnetic
Reynolds number.

However, MHD simulations suggest the critical magnetic Reynolds number
may be much higher.  For higher critical magnetic Reynolds numbers,
$Re_{\rm M,crit}\gtrsim 100$ we have found an analytical fit to the
active surface density (equations~\ref{fit} and~\ref{fit2}) that will
be a useful approximation in future time-dependent calculations.  The
dead zone structure is very sensitive to the value of the critical
magnetic Reynolds number, as seen in Fig.~\ref{layer}, but the value
is still uncertain and needs to be clarified in further work. However,
this is complicated by the fact that the $\alpha$ parameter in the
viscosity is also uncertain.

The metallicity variation between our galaxy, the LMC and the SMC is
not significant enough to affect the expected size of the dead zone in
a disc (ignoring possible differences in dust abundances).  When
comparing accretion discs in our galaxy with those in the LMC and SMC
for example, the high metallicity limit will be appropriate. In order
to explain the observed difference in disc lifetimes with metallicity
\citep[e.g.][]{demarchi10,demarchi11} some other effect must be taken
into account.

\section*{Acknowledgements}

We acknowledge useful comments from the anonymous referee. RGM thanks
the Space Telescope Science Institute for a Giacconi Fellowship.  SHL
acknowledges support from NASA grant NNX07AI72G. JEP thanks the
Collaborative Visitor Program at STScI for its support and
hospitality.

%\clearpage

\label{lastpage}
\end{document}